\documentclass[sigconf]{acmart}
\usepackage{tabularx}
\usepackage{array}  
\usepackage{booktabs}  
\usepackage{amsmath}   
\usepackage{tabularx}  
\usepackage{graphicx}  
\usepackage{array}  
\usepackage{multirow} 
\usepackage{xspace}
\usepackage{booktabs}
\usepackage{tcolorbox}
\usepackage{listings}%
\usepackage{subfig}

\begin{document}

\title{EvoReason: Self-Evolving Reasoning Primitive-Guided On-Policy Distillation for Latent Reasoning in Generative Recommendation}
\author{Zhuang Zhuang}
\affiliation{%
  \institution{Kuaishou Technology}
  \city{Beijing}
  \country{China}
}
\email{zhuangzhuang@kuaishou.com}

\author{Zhipeng Wei}
\affiliation{%
  \institution{Kuaishou Technology}
  \city{Beijing}
  \country{China}
}
\email{weizhipeng@kuaishou.com}

\author{Rongfeng Guo}
\affiliation{%
  \institution{Shenzhen University}
  \city{Shenzhen}
  \country{China}
}
\email{2023096062@email.szu.edu.cn}

\author{Shijie Li}
\affiliation{%
  \institution{Kuaishou Technology}
  \city{Beijing}
  \country{China}
}
\email{lishijie05@kuaishou.com}

\author{Peng Zhao}
\affiliation{%
  \institution{Kuaishou Technology}
  \city{Beijing}
  \country{China}
}
\email{zhaopeng14@kuaishou.com}

\author{Jie Chen}
\authornote{Corresponding author.}
\affiliation{%
  \institution{Kuaishou Technology}
  \city{Beijing}
  \country{China}
}
\email{chenjie20@kuaishou.com}

\author{Fei Pan}
\authornotemark[1]
\affiliation{%
  \institution{Kuaishou Technology}
  \city{Beijing}
  \country{China}
}
\email{panfei05@kuaishou.com}


\begin{abstract}
Generative recommendation benefits from reasoning-enhanced inference, and latent reasoning offers an efficient paradigm by encoding intermediate reasoning processes into compact continuous representations for latency-sensitive deployment. Despite its efficiency, existing latent reasoning approaches typically rely on directly distilling raw chain-of-thought (CoT) trajectories into latent representations, assuming that textual reasoning traces provide sufficient supervision. However, recommendation reasoning trajectories contain diverse reasoning processes with redundant expressions and unstable reasoning paths, making raw CoT supervision suboptimal for learning transferable latent reasoning representations.

To address this challenge, we propose \textbf{EvoReason}, a self-evolving latent reasoning framework that adaptively aligns explicit reasoning supervision with the student’s latent reasoning space through primitive-guided on-policy distillation. 
First, EvoReason extracts reusable reasoning primitives from high-quality agentic recommendation trajectories, where each primitive captures an essential reasoning behavior and serves as a pseudo-tool for structured teacher reasoning. 
Then, based on these primitives, we equip the teacher with primitive-aware reasoning capabilities, enabling it to generate structured CoT supervision with reduced redundancy and improved consistency. 
Finally, during latent reasoning optimization, EvoReason introduces a self-evolving on-policy distillation mechanism, where the primitive-guided teacher refines the student's on-policy reasoning behaviors while the primitive-guided reasoning process evolves according to the student's latent reasoning outcomes. 
Through this closed-loop co-evolution, policy updates continuously improve latent reasoning behaviors while primitive-guided reasoning generation is refined according to the resulting latent reasoning outcomes, enabling progressively better-aligned CoT supervision and more effective reasoning transfer through on-policy distillation and latent alignment while maintaining efficient inference without explicit CoT generation.
Extensive experiments on multiple recommendation benchmarks demonstrate that EvoReason consistently outperforms existing generative and latent reasoning methods, validating the effectiveness of primitive-guided reasoning supervision.
\end{abstract}




\maketitle

\begin{figure}[!t]
    \centering
    \includegraphics[width=1\columnwidth]{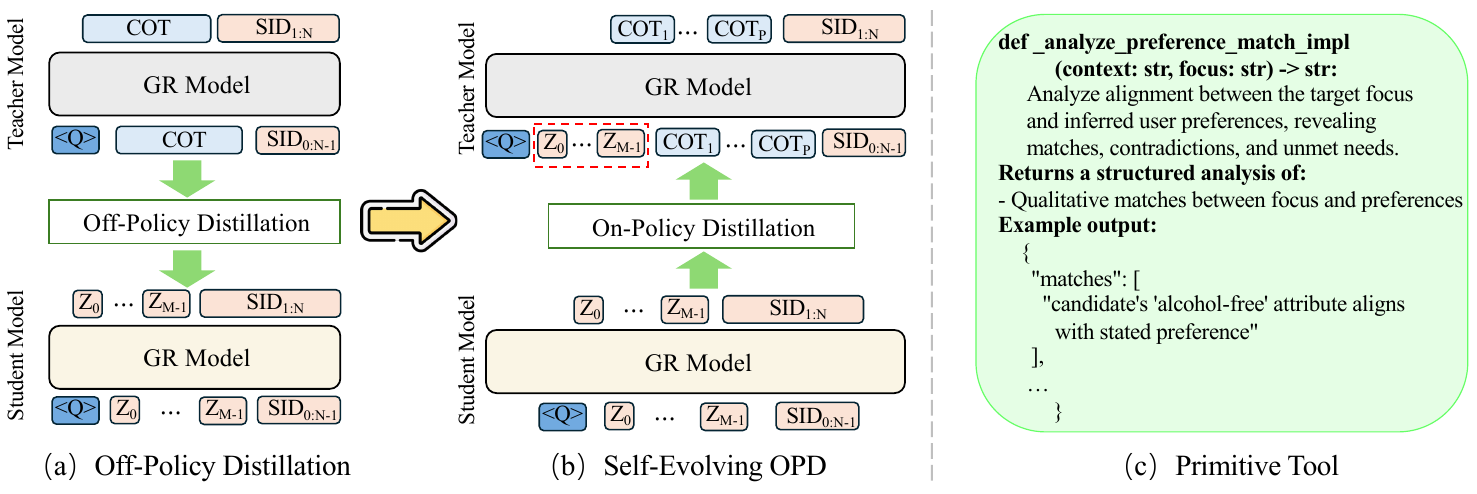}
    \caption{Framework comparison between (1) conventional off-policy distillation and (2) the proposed self-evolving OPD framework, which enables the teacher to leverage multiple reasoning primitives to generate structured CoT supervision from the student’s latent reasoning states and distill the refined reasoning back to the student. (3) an illustrative example of function-like reasoning primitives for structured explicit reasoning generation. Each primitive resembles a Python-style function, consisting of input parameters, structured reasoning procedures, and output format constraints that jointly regulate the generation of explicit reasoning trajectories.}
    \label{fig:evo}
\end{figure}
\section{Introduction}
The rapid development of large language models (LLMs) has reshaped the trajectory of recommender systems and introduced new opportunities for generative recommendation~\cite{zhuang2026sinkrec}. It enables a paradigm shift from traditional candidate retrieval~\cite{zhuang2024tau, zhuang2025mgstdn} and ranking to autoregressive generation of target item identifiers~\cite{hecomprehensive}. Meanwhile, the emerging reasoning capability of LLMs has become a major research focus, as demonstrated by models such as DeepSeek-R1~\cite{guo2025deepseek} and OpenAI-o1, where chain-of-thought (CoT) reasoning has shown significant potential for complex task analysis~\cite{xu2025toward}. These advances have further promoted the application of LLM reasoning in recommender systems~\cite{team2026onereason,zheng2026recgpt}. The core advantage of LLM-based reasoning lies in leveraging world knowledge to infer users’ complex latent preferences and behavioral patterns from historical interactions, which are difficult to capture with conventional sequential models~\cite{tsai2024leveraging,bismay2025reasoningrec}. By aligning the strengths of LLM reasoning with the limitations of traditional recommendation approaches, this paradigm opens new research directions for personalized recommender systems.

Existing LLM reasoning approaches for recommendation can be broadly categorized into explicit reasoning and latent reasoning paradigms. Explicit reasoning methods follow the CoT paradigm of LLMs and enhance recommendation through textual reasoning supervision. For example, OneRec-Think~\cite{liu2025onerec}, ReasoningRec~\cite{bismay2025reasoningrec}, $R^2ec$~\cite{you2026r}, and Think2Go~\cite{zhuang2026think2go} introduce explicit reasoning capabilities via supervised fine-tuning, while REG4Rec~\cite{xing2025reg4rec} and R4ec~\cite{gu2025r} improve reasoning through reflection and refinement mechanisms. Recent agent-based methods, such as AgentCF~\cite{zhang2024agentcf}, MemRec~\cite{chen2026memrec}, and RecThinker~\cite{zhang2026recthinker}, further incorporate collaborative signals, memory, and external tools to enrich the reasoning process. However, generating lengthy reasoning chains introduces considerable inference latency, limiting their practical deployment.
Latent reasoning methods address this issue by encoding reasoning processes into compact hidden representations, which provide higher information density and more efficient inference. Existing studies explore latent reasoning from different perspectives: LARES~\cite{payne2008lares} and ReaRec~\cite{tang2026think} model evolving preferences through recurrent structures and reasoning position embeddings, LatentR3~\cite{zhang2025reinforced} optimizes latent tokens with reinforcement learning, and FLR~\cite{gao2026factorized} decomposes latent reasoning into disentangled preference factors. RecGPT-V3~\cite{zheng2026recgpt} and LASAR~\cite{chen2026lasar} further distill CoT supervision into latent tokens. Nevertheless, existing latent reasoning methods mainly rely on static reasoning supervision and lack a mechanism to continuously evolve reasoning knowledge and improve latent reasoning through interaction with explicit reasoning.

Although existing latent reasoning approaches have achieved promising progress, they still suffer from several fundamental limitations that hinder their effectiveness:
\begin{itemize}
    \item \textbf{Inefficient Reasoning Transfer from Explicit CoT to Latent Tokens.}
            Existing methods leverage explicit CoT trajectories as semantic supervision to train latent reasoning representations. However, directly distilling full CoT sequences requires latent tokens to learn both high-level reasoning behaviors and their instance-specific linguistic realizations. Without identifying reusable reasoning structures behind these trajectories, such supervision entangles essential reasoning patterns with redundant variations, resulting in inefficient transfer to compact latent representations.
    \item \textbf{Static Reasoning Supervision.} 
            Existing approaches typically generate teacher reasoning trajectories offline and keep them unchanged during latent reasoning optimization. However, the latent reasoner continuously evolves during training, while fixed teacher demonstrations cannot adapt to the student's changing reasoning capability, leading to suboptimal supervision and limited reasoning transfer.
    \item \textbf{Lack of Controllable Explicit Reasoning.}
            Existing methods typically rely on free-form CoT generation, where reasoning trajectories are generated without explicit guidance over reasoning behaviors. Such unconstrained generation captures diverse and inconsistent reasoning processes, making it difficult to maintain stable reasoning patterns across different samples. Without controllable structures for planning, refinement, and decision-making behaviors, the generated supervision provides limited guidance for learning transferable latent reasoning capabilities.         
\end{itemize}

To this end, we propose EvoReason, a primitive-guided latent reasoning framework with self-evolving on-policy distillation for generative recommendation.
\textbf{To address inefficient reasoning transfer from explicit CoT to latent tokens}, EvoReason extracts reusable reasoning primitives from high-quality agentic recommendation trajectories. A ReAct-style teacher generates trajectories containing intermediate decision-making behaviors beyond conventional CoT texts, from which recurring reasoning patterns are abstracted into a primitive library. These primitives provide compact and transferable semantic supervision, allowing latent tokens to focus on essential reasoning behaviors rather than redundant linguistic patterns.
\textbf{To overcome static reasoning supervision}, EvoReason introduces a self-evolving on-policy distillation framework. During training, the latent reasoner generates its own on-policy behaviors, and the primitive-guided teacher refines these behaviors into adaptive reasoning trajectories. The primitive library is continuously updated based on latent reasoning outcomes, enabling teacher supervision to evolve with the student's changing reasoning capability instead of relying on fixed demonstrations.
\textbf{To alleviate the lack of reasoning trajectory constraints}, EvoReason uses the primitive library as structured guidance for teacher reasoning generation. By organizing reasoning processes into reusable pseudo-tools, the teacher produces controllable and consistent reasoning trajectories with reduced redundant variations, providing stable reasoning supervision for latent reasoning learning.

Our contributions are summarized as follows:
\begin{itemize}
\item We propose EvoReason, a primitive-guided latent reasoning framework for generative recommendation that constructs a self-evolving reasoning supervision loop, where primitive-guided teacher reasoning is continuously refined according to student feedback.

\item We introduce a self-evolving on-policy distillation mechanism that leverages analyzer-driven primitive evolution to generate adaptive CoT supervision, enabling latent tokens to learn more effective reasoning representations.

\item Extensive experiments on multiple recommendation benchmarks demonstrate that EvoReason consistently outperforms existing generative recommendation and latent reasoning methods.
\end{itemize}

\section{Related Work}
\subsection{LLM Latent Reasoning}
Latent reasoning has recently emerged as an efficient alternative to explicit Chain-of-Thought (CoT) reasoning~\cite{wei2022chain} for reducing inference latency and deployment cost. Instead of generating verbose reasoning in the discrete text space, latent reasoning performs intermediate computation directly in a continuous representation space, inheriting the observation that increasing inference-time computation can improve the reasoning capability of Transformer models~\cite{zheng2026recgpt}.
Early studies demonstrated that inserting additional latent tokens before answer prediction enables implicit multi-step reasoning~\cite{strobl2024formal}. Coconut~\cite{hao2024training} introduced latent reasoning through curriculum learning~\cite{deng2024explicit} by recursively propagating autoregressive hidden states as intermediate reasoning tokens, but remained inferior to explicit CoT due to catastrophic forgetting during curriculum transition. Subsequent approaches further distilled explicit CoT into latent representations through supervised fine-tuning~\cite{deng2023implicit}.
More recently, CoDi~\cite{shen2025codi}, KAVA~\cite{kuzina2025kava}, and CoLaR~\cite{tan2026think} independently proposed dynamically compressing explicit CoT into a flexible latent reasoning space. They supervise latent reasoning by aligning hidden states of designated reasoning tokens, compressed KV caches, or merged embeddings of consecutive reasoning tokens, respectively.
In recommendation systems, early efforts also explored increasing implicit computation to capture complex user intents. LARES~\cite{liu2025lares} and ReaRec~\cite{tang2026think} modeled implicit preference reasoning through recurrent reasoning and reasoning-position embeddings, respectively. LatentR3~\cite{zhang2025reinforced} and FLR~\cite{gao2026factorized} further incorporated reinforcement learning and factorized latent reasoning to provide stronger supervision for latent reasoning. LASAR~\cite{chen2026lasar} distilled CoT reasoning into latent tokens to alleviate the semantic grounding gap between latent reasoning and semantic IDs.

However, existing methods directly distill raw reasoning trajectories into latent representations without disentangling the reusable reasoning primitives they contain. Consequently, multiple reasoning behaviors are entangled into a single supervision signal, preventing latent reasoning from learning reusable reasoning behaviors.

\subsection{On-Policy Distillation}
On-policy distillation (OPD)~\cite{song2026survey} extends conventional knowledge distillation~\cite{hinton2015distilling} by performing teacher supervision on trajectories sampled from the current student policy, thereby aligning supervision with the student’s on-policy visitation distribution~\cite{xu2026deepseek, hou2026uni,wu2026seed}. This on-policy formulation alleviates the distribution mismatch between training and inference, while dense token-level distillation provides substantially richer supervision than trajectory-level rewards used in RL-based optimization~\cite{zhou2026turnopd}. Unlike supervised fine-tuning, which relies on fixed off-policy demonstrations and cannot exploit the student's evolving behavior, OPD continuously adapts the supervision to the student's current policy, enabling more efficient and stable optimization~\cite{zhao2026training}.
Recent studies have significantly advanced OPD from different perspectives. G-OPD~\cite{yang2026learning} establishes a unified theoretical framework by formulating OPD under KL-constrained reinforcement learning. Flow-OPD~\cite{fang2026flow} introduces multi-teacher dense supervision to alleviate sparse-reward optimization. PG-OPD~\cite{zhao2026prefix}, CCOPD~\cite{lin2026same}, ReOPD~\cite{liao2026multi}, and OPID~\cite{yang2026opid} improve the reliability of supervision by extracting reusable knowledge or constructing privileged signals from teacher behaviors, while D-OPSD~\cite{jiang2026d} extends OPD to text-to-image diffusion models for continual step-distilled training.

Despite these advances, existing OPD methods mainly distill explicit behaviors and overlook the transfer of reasoning capabilities into compact latent representations. EvoReason addresses this challenge with self-evolving primitive-guided supervision, which continuously improves the alignment between explicit reasoning knowledge and latent representations, enabling effective latent reasoning transfer without explicit CoT inference.

\section{Preliminaries}
\subsection{Problem Definition}
Let $\mathcal{U}$ and $\mathcal{V}$ denote the sets of users and items, respectively. Each user $u\in\mathcal{U}$ is associated with a chronological interaction history $V_u=(v_1,\ldots,v_n)$, where $v_i\in\mathcal{V}$. Following OneRec-Think\cite{zhou2025onerec, liu2025onerec}, each item is mapped to a semantic ID (SID) generated from multimodal information and collaborative signals. The interaction history is therefore represented as a SID sequence $S_u=(s_{v_1},\ldots,s_{v_n})$.

Given the prompted user context $x=\mathcal{P}(S_u)$, our goal is to jointly perform latent reasoning and next-item recommendation within a unified autoregressive framework. Instead of generating explicit reasoning tokens, the model first performs iterative latent reasoning over a set of $M$ latent tokens $Z=(z_1,\ldots,z_M)$, and then predicts the target SID:
\begin{equation}
\begin{aligned}
Z &\sim P(\cdot \mid x;\theta),\\
s_{v_{n+1}} &\sim P(\cdot \mid x,Z;\theta),
\end{aligned}
\end{equation}
where $\theta$ denotes the model parameters. During training, latent reasoning is supervised by primitive-guided explicit reasoning generated from a teacher model, whereas during inference only the latent reasoning process is executed without generating intermediate textual reasoning.

\subsection{On-Policy Distillation}
We consider on-policy distillation (OPD) as an effective paradigm for reasoning transfer, where the teacher provides adaptive token-level supervision according to the trajectories sampled by the current student policy. This property is particularly suitable for our setting, where the teacher generates explicit CoT reasoning trajectories based on the student's latent states, while the student aims to internalize such reasoning behaviors into compact latent representations. By optimizing on the student's own generated trajectories, OPD alleviates the distribution mismatch between training and inference and provides dense supervision for iterative reasoning improvement.

We adopt OPD for reasoning transfer, where the teacher provides token-level supervision on trajectories sampled by the student policy. Given an input context $x$, the student policy $\pi_{\theta}$ samples an on-policy trajectory $\hat{y}\sim\pi_{\theta}(\cdot|x)$, and the teacher policy $\pi_T$ provides token-level guidance on the visited states. The OPD objective is formulated as:

\begin{equation}
\mathcal{L}_{\mathrm{OPD}}
=
\mathbb{E}_{x\sim\mathcal{D},\hat{y}\sim\pi_{\theta}}
\left[
\sum_{t=1}^{T}
D_{\mathrm{KL}}
\left(
\pi_{\theta}(\cdot|s_t)
\|
\pi_T(\cdot|s_t)
\right)
\right],
\end{equation}

where $s_t=(x,\hat{y}_{<t})$ denotes the student-visited prefix at decoding step $t$. Compared with supervised fine-tuning on fixed reference trajectories, OPD performs supervision directly on the student's visitation distribution, thereby alleviating distribution mismatch while providing dense token-level learning signals throughout the generation process.

\begin{figure*}[t]
\centering
\includegraphics[width=1.0\textwidth]{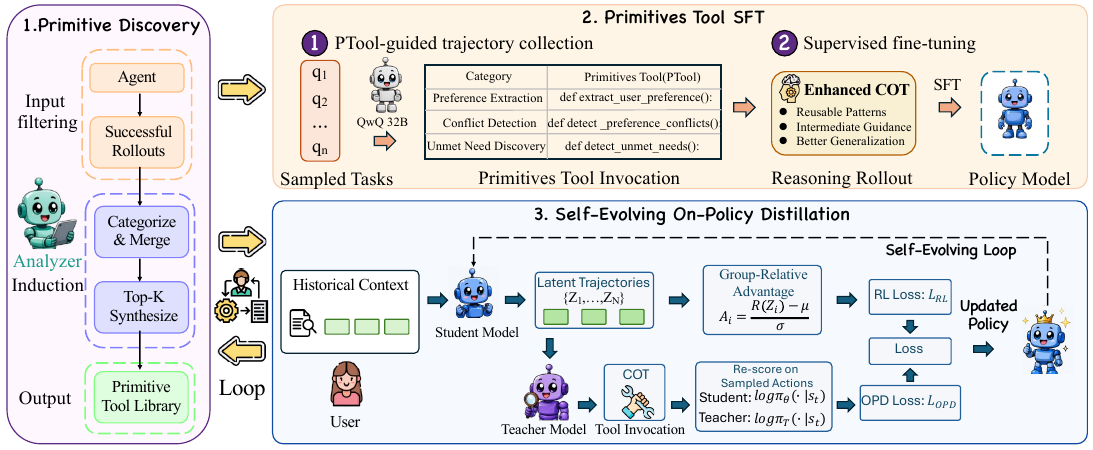} 
\caption{Overview of EvoReason. Stage 1 discovers reusable reasoning primitives from high-quality trajectories and constructs a primitive tool library for structured reasoning. Stage 2 fine-tunes the teacher to invoke primitives and generate primitive-guided CoT supervision. Stage 3 performs self-evolving on-policy latent reasoning distillation, jointly optimizing policy improvement and reasoning transfer through outcome-based RL and latent-conditioned OPD.}
\label{fig:architecture}
\end{figure*}

\section{Method}
\subsection{Overview}
We present \textbf{EvoReason}, a self-evolving primitive-guided latent reasoning framework for generative recommendation. EvoReason is motivated by the observation that effective latent reasoning distillation requires reasoning supervision aligned with the latent reasoning process, as arbitrary teacher-generated CoT trajectories may not be directly compatible with latent representations. To obtain structured and adaptive reasoning supervision, EvoReason leverages ReAct-style agentic trajectories to discover reusable reasoning primitives, which are continuously evolved through latent reasoning optimization. These evolving primitives guide teacher reasoning generation and enable more effective reasoning transfer into compact latent representations.

As shown in Figure~\ref{fig:architecture}, EvoReason consists of three stages.
In the first two stages, we discover reasoning primitives from agentic trajectories and fine-tune a primitive-aware teacher to generate structured reasoning supervision.
In the third stage, EvoReason performs self-evolving on-policy latent reasoning distillation, where the teacher refines latent behaviors and the updated reasoning experiences further evolve the primitive library.
This enables continuous reasoning transfer from explicit teacher behaviors to compact latent representations without requiring explicit CoT generation at inference.

\subsection{Self-Evolving Reasoning Primitive Discovery}
\label{sec:primitive}

\subsubsection{Agentic Reasoning Experience Collection}
To discover reusable reasoning behaviors, EvoReason first collects high-quality recommendation trajectories from an agentic teacher equipped with ReAct-style tool-use capabilities~\cite{lei2026inducing}. The teacher performs multi-step reasoning involving evidence extraction, preference analysis, intent inference, and recommendation decisions. Each trajectory records intermediate reasoning actions and outcomes, forming a reasoning experience pool:
\begin{equation}
\mathcal{D}_{reason}=\{\tau_i\}_{i=1}^{N}.
\end{equation}

Different from conventional CoT traces that only contain final textual reasoning, these agentic trajectories expose the underlying decision-making procedures, providing richer signals for reasoning primitive discovery.

\subsubsection{Reasoning Primitive Induction}

Based on the collected reasoning experiences, we identify recurring reasoning patterns and abstract them into a primitive library:
\begin{equation}
\mathcal{P}=\{p_1,p_2,\cdots,p_K\},
\end{equation}
where each primitive represents a reusable reasoning operation with a semantic description and invocation interface.

The primitive library serves as a set of pseudo-tools that guides teacher reasoning generation. Given an input context $x$, the teacher generates structured reasoning trajectories by invoking primitives:
\begin{equation}
c=G(x,\mathcal{P}),
\end{equation}
where the generated trajectory follows primitive-guided reasoning patterns rather than unconstrained CoT generation. This enables the teacher to reuse discovered reasoning behaviors and provide more structured supervision for latent reasoning distillation.

\subsubsection{Self-Evolving Primitive Library}

Instead of treating the primitive library as fixed knowledge, EvoReason continuously evolves $\mathcal{P}$ according to the optimization behaviors of the latent reasoner. During training, the primitive-guided teacher generates reasoning trajectories based on the current primitive library $\mathcal{P}^{m}$:
\begin{equation}
c^{m}=G(x,\mathcal{P}^{m}).
\end{equation}

After latent reasoning optimization, the resulting recommendation behaviors provide feedback for primitive refinement. Successful trajectories reveal effective reasoning patterns, while failed trajectories expose missing or ineffective reasoning behaviors. Therefore, new primitives are induced from the updated reasoning experiences and merged with the existing library:
\begin{equation}
\mathcal{P}^{m+1}
=
\operatorname{Update}
(\mathcal{P}^{m},\mathcal{R}_{latent}),
\end{equation}
where $\mathcal{R}_{latent}$ denotes the reasoning behaviors collected from the current latent reasoner.

Through this self-evolving process, the primitive library progressively adapts to the latent reasoner's reasoning capability, enabling the teacher to generate increasingly effective structured reasoning supervision for latent reasoning distillation.

\subsection{Primitive-Aware Teacher Supervised Fine-Tuning}
\label{sec:sft}

After obtaining the self-evolving primitive library, we perform supervised fine-tuning to initialize a primitive-aware reasoning teacher. The goal of this stage is to equip the teacher policy with primitive invocation ability, enabling it to generate structured reasoning trajectories that can provide explicit supervision for subsequent latent reasoning distillation.

\subsubsection{Primitive-Guided Reasoning Data Construction}

Given a recommendation context $x=[s,u,\mathcal{H}_u]$, the primitive library $\mathcal{P}$ serves as a set of pseudo-tools that organize the reasoning process. Instead of generating unconstrained chain-of-thought reasoning, the teacher is instructed to invoke appropriate primitives and produce structured reasoning trajectories:

\begin{equation}
c=
[
\langle p_1\rangle,c_1,
\langle p_2\rangle,c_2,\cdots,
\langle p_K\rangle,c_K
],
\end{equation}

where $\langle p_k\rangle$ denotes the invoked primitive and $c_k$ represents the corresponding reasoning content. These primitive-guided trajectories expose intermediate decision-making behaviors while maintaining a consistent reasoning structure.

Based on the generated trajectories and recommendation targets, we construct the teacher fine-tuning dataset:

\begin{equation}
\mathcal{D}_{sft}
=
\{(x_i,c_i,y_i)\}_{i=1}^{N},
\end{equation}

where $y_i$ denotes the target recommendation item.

\subsubsection{Primitive-Aware Teacher Optimization}

Using the constructed dataset, we fine-tune the teacher policy with the standard causal language modeling objective:

\begin{equation}
\mathcal{L}_{sft}
=
-\mathbb{E}_{(x,c,y)\sim\mathcal{D}_{sft}}
\sum_t
\log p_{\phi}(Y_t|x,Y_{<t}),
\end{equation}

where $Y$ consists of primitive invocation tokens, structured reasoning tokens, and recommendation tokens.

After fine-tuning, the teacher policy learns to generate primitive-guided reasoning trajectories:

\begin{equation}
(c,y)=G_{\phi}(x,\mathcal{P}),
\end{equation}

which serve as adaptive explicit reasoning supervision for the following self-evolving on-policy latent reasoning distillation stage.

\subsection{Self-Evolving On-Policy Latent Reasoning Distillation}
\label{sec:opd}
The third stage performs self-evolving on-policy latent reasoning distillation, where the primitive library, policy optimization, and reasoning distillation jointly evolve. The student improves latent reasoning through outcome-based reinforcement learning, while a primitive-guided teacher refines these behaviors into explicit reasoning trajectories. The resulting reasoning knowledge is transferred back through confidence-gated OPD and latent KV alignment, forming a self-evolving optimization loop.

\subsubsection{On-policy Latent Policy Optimization}

At each iteration, the student first performs latent reasoning refinement
to produce its current recommendation behavior~\cite{shen2025codi}. Given the
recommendation context $x=[s,u,\mathcal{H}_u]$, the latent states are
initialized by primitive-aware slots:

\begin{equation}
\mathbf{z}^{(0)}=\mathrm{Proj}(\mathbf{E}_{slot}),
\end{equation}

and iteratively refined as:

\begin{equation}
\mathbf{z}^{(t)}
=
\mathrm{Proj}
(
f_{\theta}
([\mathbf{x}_{enc};\mathbf{z}^{(t-1)};\mathbf{x}_{eot}])
_{[enc:enc+M]}
),
\quad t=1,\ldots,T .
\end{equation}

The refined latent states are then used to generate the on-policy
recommendation trajectory:

\begin{equation}
y_o=f_{\theta}
([\mathbf{x}_{enc};\mathbf{z}^{(T)};\mathbf{x}_{eot}]).
\end{equation}

To improve the latent reasoning policy, we optimize the student using
outcome-driven reinforcement learning on its own generated trajectories.
For each context $x$, the frozen policy snapshot
$\pi_{\theta_{\mathrm{old}}}$ samples a trajectory group
$G_x=\{\tau_1,\ldots,\tau_N\}$, where
$\tau_i\sim\pi_{\theta_{\mathrm{old}}}(\cdot|x)$.
The trajectory reward is defined by the token-level SID matching accuracy:

\begin{equation}
R(\tau_i)=
\sum_j\mathbb{I}(y_{i,j}=y_j^*).
\end{equation}

Following GRPO, we compute the group-relative advantage
$A_i=(R(\tau_i)-\mu_G)/(\sigma_G+\epsilon)$ and assign it to valid
response tokens as $A_{i,t}=A_i m_{i,t}$. The policy is then optimized with
the clipped group-relative objective:

\begin{equation}
\mathcal{L}_{RL}
=
-\mathbb{E}
\left[
\min
\left(
\rho_{i,t}A_{i,t},
\mathrm{clip}(\rho_{i,t},1-\epsilon,1+\epsilon)A_{i,t}
\right)
\right]
+
\alpha_{KL}D_{KL}(\pi_\theta||\pi_{ref}),
\end{equation}

where $\rho_{i,t}=\exp(\ell^\theta_{i,t}-\ell^{old}_{i,t})$ denotes the
token-level importance ratio between the current policy and the rollout
policy.

The optimized latent policy continuously improves its own reasoning
behaviors, which are subsequently refined by the primitive-guided teacher
and distilled through the OPD objective.

\subsubsection{Latent-conditioned Teacher Refinement}
The teacher shares the same backbone with the student and performs a stop-gradient forward pass, providing training-time supervision without
introducing additional parameters. Directly distilling the student's
on-policy trajectory is insufficient, since early reasoning errors can
propagate through subsequent generation and result in prefix failures~\cite{jiang2026trajectory}. To address this issue, we introduce a latent-conditioned teacher that performs interleaved reasoning refinement over the student's latent behavior before distillation.

Specifically, the teacher receives the latent states and the generated
trajectory:

\begin{equation}
\mathbf{x}_{T}
=
[\mathbf{x}_{enc};
\mathrm{sg}(\mathbf{z}^{(T)});
y_o;
\mathbf{x}_{refine}],
\end{equation}

where $\mathrm{sg}(\cdot)$ denotes stop-gradient. With the evolving
primitive library $\mathcal{P}$ as structured reasoning guidance, the
teacher refines the current trajectory:

\begin{equation}
y_r
=
G_{\theta}(\mathbf{x}_{T},\mathcal{P}),
\end{equation}

where $y_r$ provides a corrected reasoning trajectory that preserves
useful on-policy behaviors while resolving unreliable reasoning patterns.
The refined trajectory is then used as the teacher signal for subsequent
confidence-gated OPD and latent alignment.

\subsubsection{On-policy Distillation and Latent Alignment}

The primitive-guided teacher trajectory is distilled back into the student
through an on-policy objective. Different from conventional distillation
with an independent teacher model, our teacher and student share the same
backbone, while the teacher provides additional reasoning context during
training. Specifically, given the student's latent trajectory, the student
evaluates the sampled recommendation tokens under the original latent
context:

\begin{equation}
\ell_t^{S}
=
\log
\pi_{\theta}
(y_t|x,\mathbf{z}^{(T)},y_{<t}),
\end{equation}

while the teacher branch evaluates the same tokens with the refined
reasoning trajectory:

\begin{equation}
\ell_t^{T}
=
\log
\pi_{\theta}
(y_t|x,\mathbf{z}^{(T)},y_r,y_{<t}).
\end{equation}

Here, the teacher signal is generated only through context augmentation,
without introducing additional parameters. To prevent unreliable teacher
signals from dominating optimization, we measure the teacher-induced
confidence gain:

\begin{equation}
g_t
=
\sigma
\left(
\beta
\,
\mathrm{sg}
(\ell_t^{T}-\ell_t^{S})
\right),
\end{equation}

where $\beta$ controls the sharpness of the confidence gate and
$\mathrm{sg}(\cdot)$ denotes stop-gradient. The resulting confidence-gated
OPD objective is:

\begin{equation}
\mathcal{L}_{OPD}
=
-
\frac{1}{\sum_t g_t}
\sum_t
g_t
\,
\mathrm{sg}(\ell_t^{T})
\log
\pi_{\theta}(y_t|x,\mathbf{z}^{(T)},y_{<t}).
\end{equation}

This objective encourages the latent policy to internalize the behavioral
improvements induced by primitive-guided reasoning while avoiding
uncertain teacher supervision.

Although OPD transfers reasoning behaviors at the token level, the explicit
teacher trajectory is not directly available during latent inference.
Therefore, we further align the teacher reasoning states with the student's
latent states. Specifically, we compress the teacher reasoning trajectory
using reasoning-aware KV selection (i.e., R-KV~\cite{cai2026r}), which retains informative
reasoning tokens from different primitive segments. The compressed teacher
KV representations are then aligned with the student's latent KV states:

\begin{equation}
\mathcal{L}_{KV}
=
\frac{1}{L}
\sum_l
(
\|K_T^l-K_S^l\|_1
+
\|V_T^l-V_S^l\|_1
),
\end{equation}

where $K_T^l,V_T^l$ denote the compressed teacher KV representations and
$K_S^l,V_S^l$ denote the latent KV states produced by the student.

The overall optimization objective combines outcome-guided reinforcement
learning with reasoning distillation and latent alignment:

\begin{equation}
\mathcal{L}
=
\mathcal{L}_{RL}
+
\lambda_{OPD}\mathcal{L}_{OPD}
+
\lambda_{KV}\mathcal{L}_{KV},
\end{equation}

where $\mathcal{L}_{RL}$ denotes the on-policy optimization objective
described above. After each update, the improved latent policy generates
new behaviors, which are further analyzed to evolve the primitive library,
closing the self-evolving reasoning loop.

\begin{table}[t]
\centering
\caption{Statistics of the Public datasets.}
\label{tab:dataset_statistics}
\small
\begin{tabular}{lccccc}
\toprule
Dataset & Users & Items & Interactions & Sparsity & Test Samples \\
\midrule
Beauty & 22,363 & 12,101 & 194,687 & 99.928\% & 22,363 \\
Sports & 35,598 & 18,357 & 294,488 & 99.955\% & 35,598 \\
\bottomrule
\end{tabular}
\end{table}

\begin{table*}[t]
\setlength{\abovecaptionskip}{0.05cm}
\setlength{\belowcaptionskip}{0.2cm}
\caption{Overall performance comparison between the baselines and EvoReason on three datasets. The bold results highlight the best results, while the second-best ones are underlined.}

\setlength{\tabcolsep}{2mm}
\resizebox{\textwidth}{!}{
\begin{tabular}{c|cccc|cccc|cccc}
\toprule
\multirow{2}{*}{\textbf{Method}} 
& \multicolumn{4}{c|}{\textbf{Beauty}}
& \multicolumn{4}{c|}{\textbf{Sports}}
& \multicolumn{4}{c}{\textbf{Industrial}}\\
\cline{2-13}

& R@5 & R@10 & N@5 & N@10
& R@5 & R@10 & N@5 & N@10
& R@5 & R@10 & N@5 & N@10\\

\midrule

Mamba4Rec
&0.0300&0.0404&0.0222&0.0255
&0.0119&0.0158&0.0084&0.0097
&0.0404&0.0902&0.0305&0.0358\\

TiM4Rec
&0.0334&0.0480&0.0212&0.0259
&0.0153&0.0219&0.0099&0.0121
&0.0420&0.0991&0.0249&0.0426\\

GRU4Rec
&0.0395&0.0584&0.0265&0.0326
&0.0190&0.0365&0.0122&0.0206
&0.0332&0.0753&0.0241&0.0403\\

SASRec
&0.0402&0.0607&0.0254&0.0320
&0.0293&0.0453&0.0176&0.0228
&0.0387&0.0869&0.0236&0.0409\\

TIGER
&0.0405&0.0623&0.0267&0.0337
&0.0288&0.0357&0.0203&0.0302
&0.0431&0.1038&0.0277&0.0465\\

HSTU
&0.0424&0.0652&0.0280&0.0353
&0.0354&0.0463&0.0233&0.0355
&0.0446&0.1010&0.0278&0.0472\\

ReaRec
&0.0450&0.0704&0.0262&0.0344
&0.0331&0.0455&0.0216&0.0352
&0.0472&0.1078&0.0288&0.0489\\

Onerec-think
&0.0563&0.0791&0.0398&0.0471
&0.0548&0.0588&0.0506&0.0519
&0.0473&0.1086&0.0308&0.0514\\

FLR
&0.0565&0.0781&0.0396&0.0466
&0.0558&0.0604&0.0512&0.0527
&0.0456&0.1044&0.0294&0.0491\\

LatentR3
&0.0570&0.0776&0.0397&0.0464
&0.0555&0.0598&0.0510&0.0524
&0.0467&0.1070&0.0304&0.0506\\

LASAR
&\underline{0.0613}&\underline{0.0829}&\underline{0.0437}&\underline{0.0507}
&\underline{0.0561}&\underline{0.0604}&\underline{0.0513}&\underline{0.0527}
&\underline{0.0495}&\underline{0.1122}&\underline{0.0313}&\underline{0.0527}\\
\midrule
EvoReason
&\textbf{0.0724}&\textbf{0.0919}&\textbf{0.0557}&\textbf{0.0601}
&\textbf{0.0657}&\textbf{0.0671}&\textbf{0.0597}&\textbf{0.0602}
&\textbf{0.0571}&\textbf{0.1260}&\textbf{0.0358}&\textbf{0.0573}\\

Improv. & +17.9\% & +10.8\% & +27.4\% & +18.5\% & +17.1\% & +11.1\% & +16.4\% & +14.2\% & +15.4\% & +12.3\% & +14.4\% & +8.7\% \\
\bottomrule
\end{tabular}
}
\label{tab:MainTable}
\end{table*}
\section{Experiment}\label{sec:exp}
\subsection{Experimental Settings}\label{sec:exp_setting}
\paragraph{Datasets.}
As shown in Table~\ref{tab:dataset_statistics}, we evaluate EvoReason on two widely used public sequential recommendation benchmarks from the Amazon Product Review dataset, namely Beauty and Sports, together with one proprietary industrial dataset. Following OneRec-Think, we adopt the same data preprocessing protocol and the leave-one-out evaluation setting. Both datasets are highly sparse, with sparsity ranging from 99.928\% to 99.955\%.

\paragraph{Baselines.}
We compare EvoReason against three groups of competitive baselines: (1) Classic sequential methods like GRU4Rec~\cite{hidasi2015session}, SASRec~\cite{kang2018self}, Mamba4Rec~\cite{liu2024mamba4rec} and TiM4Rec~\cite{fan2025tim4rec}; and (2) Generative Recommender Models, such as HSTU~\cite{zhai2024actions}, TIGER~\cite{rajput2023recommender}, ReaRec~\cite{tang2026think}; and (3) LLM-based Models, such as Onerec-think~\cite{liu2025onerec}, FLR~\cite{gao2026factorized}, LatentR3~\cite{zhang2025reinforced} and LASAE~\cite{chen2026lasar}.

\paragraph{Evaluation Metrics.}
Following OneRec-Think, we evaluate the recommendation performance of EvoReason using two widely adopted ranking metrics, namely Recall@$K$ and NDCG@$K$, with $K\in\{5,10\}$.
Recall@$K$ measures whether the ground-truth target item appears in the top-$K$ recommended results. NDCG@$K$ further considers the ranking position of the correctly recommended item, assigning higher scores when the target item appears earlier in the recommendation list.

\paragraph{Implementation Details.}
Our implementation follows the experimental protocol of OneRec-Think for a fair comparison. We adopt the OneRec-Think supervised fine-tuned Qwen3-1.7B as the backbone model. Following OneRec-Think, each item is represented by a four-level Semantic ID (SID) hierarchy, with 256 tokens at each level. We additionally introduce special tokens to indicate the beginning of explicit reasoning, latent reasoning, and primitive representations (pseudo-tools). All generative recommendation methods share the same SID encoding, training data, and prompt templates. The reasoning traces used for self-evolving primitive induction are generated by QwQ-32B. During inference, beam search with a beam size of 10 is used for recommendation generation. All experiments are conducted on a server equipped with eight NVIDIA A800 (80GB) GPUs. Furthermore, we follow the prompt template designs of OneRec-Think\footnote{\url{https://github.com/wangshy31/OneRec-Think}} and Induced\footnote{\url{https://github.com/lexilei/reasoning-primitives}} for constructing reasoning prompts.

\subsection{Overall Performance}\label{sec:overall_performance}
Table~\ref{tab:MainTable} summarizes the overall performance of all compared methods on the three datasets. From the results, we draw the following observations.
\textbf{(1) Reasoning-based recommenders consistently outperform conventional sequential and generative recommendation models.} Compared with generative recommenders, they incorporate explicit reasoning to better capture user preferences and behavioral patterns. Compared with sequential models, they further leverage LLM world knowledge to infer latent user intent beyond sequential dependencies, leading to more accurate recommendations.
\textbf{(2) Dense CoT supervision is substantially more effective than alternative supervision signals for latent reasoning.} Latent reasoning optimization remains challenging without explicit semantic guidance. Existing methods rely on latent objectives (ReaRec), structural regularization (FLR), or reinforcement learning rewards (LatentR3), which provide limited semantic signals or sparse feedback. In contrast, CoT-based token-level supervision offers dense semantic guidance and fine-grained credit assignment, leading to more effective latent reasoning optimization.
\textbf{(3) Self-evolving distillation is more effective than static CoT distillation.} Unlike LASAR, which relies on fixed reasoning trajectories, EvoReason continuously evolves the primitive library according to the student's latent reasoning behaviors. This enables the teacher to generate adaptive CoT supervision that better matches the latent reasoning process, resulting in more effective reasoning transfer.

\subsection{Ablation Study}
Table~\ref{tab:ablation} reports the ablation results on the Beauty dataset. Our framework consists of two key components: (1) the self-evolving reasoning primitive library and (2) OPD-based reasoning distillation from explicit CoT to latent reasoning. To evaluate their individual contributions, we remove each component separately and draw the following observations.
\textbf{(1) w/o Primitive:} Removing the primitive library eliminates the structural constraints on CoT generation and results in less structured reasoning trajectories. Although this variant still outperforms LASAR due to the on-policy optimization of OPD, its performance drops noticeably. This demonstrates that organizing CoT into reusable reasoning primitives provides more structured and effective supervision for latent reasoning.
\textbf{(2) w/o Evolve:} Disabling primitive evolution keeps the primitive library fixed throughout training. Although a static library can still capture common reasoning patterns, it cannot adapt to the evolving latent reasoning policy. In contrast, the self-evolving mechanism updates primitives based on the student's latest reasoning behaviors, enabling more adaptive CoT supervision for latent reasoning distillation.
\textbf{(3) w/o OPD:} Replacing OPD with conventional knowledge distillation removes the on-policy teacher-student interaction and relies on static supervision. As a result, the teacher no longer provides guidance on the trajectories visited by the student, leading to larger training-inference distribution mismatch and weaker reasoning transfer.

\begin{table}[t]
\setlength{\abovecaptionskip}{0.05cm}
\setlength{\belowcaptionskip}{0.2cm}
\centering
\caption{Ablation Study of different variants of EvoReason on Beauty dataset.}
\begin{tabular}{ccccc}
\toprule
\textbf{Training Method}  & \textbf{R@5} & \textbf{R@10} & \textbf{N@5} & \textbf{N@10}  \\
\midrule
w/o Primitive & 0.0636 & 0.0841 & 0.0463 & 0.0533 \\
w/o OPD  & 0.0689 & 0.0883 & 0.0502 & 0.0574 \\
w/o Evolve  & 0.0675 & 0.0877 & 0.0493 & 0.0527 \\
\midrule
EvoReason  & \textbf{0.0724} & \textbf{0.0919} & \textbf{0.0557} & \textbf{0.0601} \\
\bottomrule
\end{tabular}
\label{tab:ablation}
\end{table}

\subsection{Online Results}

We conducted an online A/B test in a production advertising system to evaluate EvoReason. LASAR, the production recall baseline, served as the control, while EvoReason replaced LASAR in the treatment bucket. The experiment ran for one week, with 20\% of live traffic allocated to the treatment bucket. Except for the recall model, the two buckets used the same downstream ranking and bidding models and the same serving configuration. Table~\ref{tab:evoreason_results} reports the relative changes of EvoReason over LASAR on business and recall-pathway metrics.

\textbf{Business metrics.}
EvoReason improves ADVV (Advertiser Value) by $8.11\%$ and platform revenue by $6.23\%$. Since the recall model is the only experimental variable, the observed gains can be attributed to the candidate set produced by EvoReason under the same downstream pipeline.

\textbf{Recall-pathway metrics.}
EvoReason increases the show ratio by $4.56\%$ and the average watch time of displayed ads by $5.39\%$. The higher show ratio indicates that a larger fraction of the recalled candidates passes downstream ranking and filtering, while the watch-time gain indicates that the displayed ads elicit stronger user engagement. Together, the pathway-level and business-level results suggest that the improvement originates from the retrieval stage rather than from a redistribution of traffic across candidates.

\begin{table}[t]
\centering
\caption{Online A/B Test Results of EvoReason}
\label{tab:evoreason_results}
\resizebox{1.0\linewidth}{!}{
\begin{tabular}{lcccc}
\toprule
\multirow{2}{*}{Method} 
& \multicolumn{2}{c}{Business Metrics} 
& \multicolumn{2}{c}{Recall Pathway Metrics} \\
\cmidrule(lr){2-3} \cmidrule(lr){4-5}
& ADVV & Revenue & Show Ratio & Avg Watch Time \\
\midrule
EvoReason & +8.11\% & +6.23\% & +4.56\% & +5.39\% \\
\bottomrule
\end{tabular}
}
\end{table}

\begin{figure}[!t]
    \centering
    \includegraphics[width=1\columnwidth]{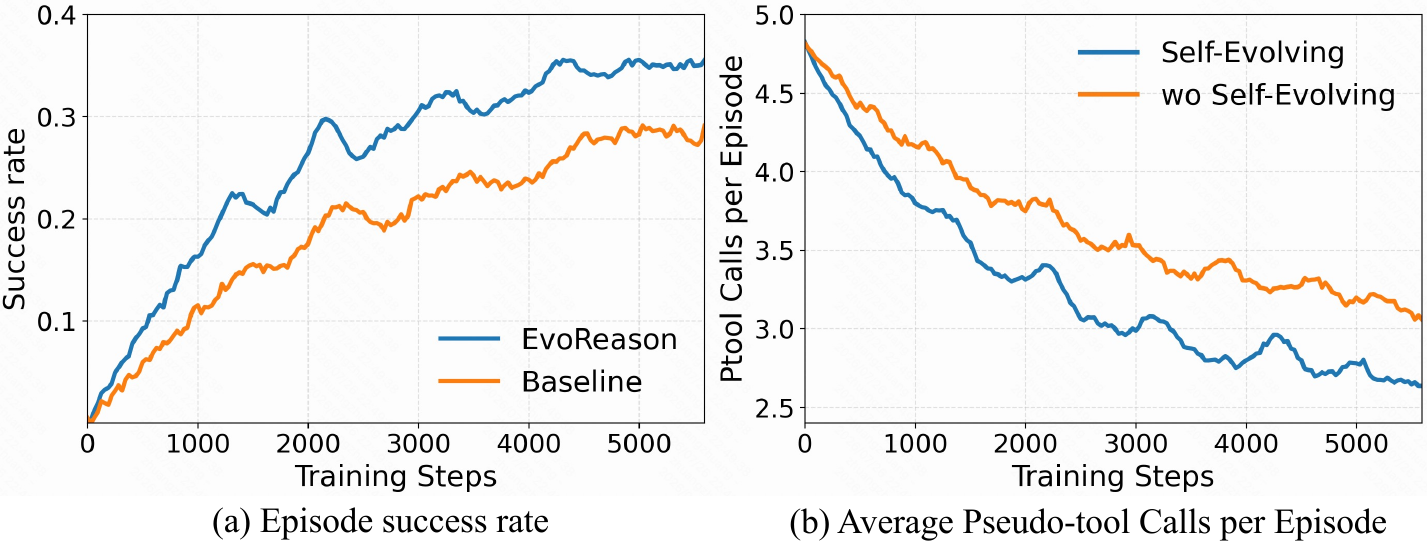}
    \caption{Training Dynamics of Self-Evolution on the Beauty Dataset. (a) Recommendation success rate comparison between EvoReason and LASAR. (b) Primitive invocation frequency comparison between self-evolving and non-evolving variants.}
    \label{fig:evo}
\end{figure}

\subsection{Self-Evolution Analysis}
To further analyze the benefits of self-evolution, we investigate the evolution of the student's recommendation success rate and the number of primitive invocations during training. In Figure~\ref{fig:evo}(a), we compare the success rate progression of EvoReason and LASAR, where both methods distill explicit reasoning into latent reasoning representations. The success rate curves diverge at an early stage: EvoReason reaches 0.21 at around 1.4K training steps, while LASAR remains at approximately 0.15. Moreover, this performance gap is consistently maintained throughout training. This improvement can be attributed to the self-evolving primitive library, which provides more effective CoT supervision for latent tokens and enables more efficient reasoning in the latent space.

Figure~\ref{fig:evo}(b) compares the number of primitive invocations between EvoReason with and without the self-evolution mechanism. We observe that the self-evolving variant reduces the average primitive invocation frequency more rapidly during training. This is because the evolving primitive library continuously refines the reasoning patterns and provides more concise and effective primitives for latent reasoning distillation, reducing unnecessary primitive usage during the reasoning process.

\begin{table}[t]
\centering
\caption{Cold-start recommendation performance on Beauty and Sports.}
\label{tab:cold_start}
\setlength{\tabcolsep}{6pt}
\begin{tabular}{llcc}
\toprule
\textbf{User Group} & \textbf{Model} &
\textbf{Beauty} &
\textbf{Sports} \\
\cmidrule(lr){3-3}\cmidrule(lr){4-4}
 & & \textbf{Recall@5} & \textbf{Recall@5} \\
\midrule

Inactive    & LASAR     & 0.0542 & 0.0493 \\
Normal      & LASAR     & 0.0623 & 0.0569 \\
Very Active & LASAR     & 0.0635 & 0.0574 \\

\midrule

Inactive    & EvoReason & 0.0684 & 0.0607 \\
Normal      & EvoReason & 0.0743 & 0.0672 \\
Very Active & EvoReason & 0.0762 & 0.0693 \\

\bottomrule
\end{tabular}
\end{table}

\subsection{User Cold-start Analysis}
To better understand how reasoning alleviates the cold-start problem, we categorize users into three groups based on the number of training trajectories: the bottom 30\% as \emph{inactive}, the top 30\% as \emph{very active}, and the remaining users as \emph{normal}. We compare our method against the state-of-the-art baseline LASAR, with the results reported in Table~\ref{tab:cold_start}. Our method consistently outperforms LASAR across all groups, with the largest improvements observed for inactive users. These results suggest that our self-evolving explicit reasoning together with implicit reasoning-conditioned OPD provides richer supervision for learning latent reasoning, enabling the model to better capture user preferences and behavioral patterns. Consequently, the semantic reasoning capability of LLMs is more effectively exploited to mitigate the cold-start problem.

\section{Conclusion}
In this work, we propose EvoReason, a primitive-guided latent reasoning framework with self-evolving on-policy distillation for generative recommendation. EvoReason addresses the challenge of transferring explicit reasoning capabilities into compact latent representations by leveraging a self-evolving primitive-guided teacher-student framework. Specifically, EvoReason first discovers reusable reasoning primitives from high-quality agentic recommendation trajectories and organizes them into a structured primitive library. These primitives act as pseudo-tools to guide the teacher in generating structured reasoning trajectories with reduced redundancy and improved consistency. During latent reasoning optimization, the primitive-guided teacher continuously refines the student's on-policy behaviors, while the primitive library evolves according to the latent reasoning outcomes. The resulting reasoning knowledge is progressively internalized into the latent policy through on-policy distillation and latent alignment, enabling adaptive and efficient reasoning transfer without relying on long reasoning chains during inference. Extensive experiments across multiple benchmarks demonstrate that EvoReason consistently improves recommendation performance, validating the effectiveness of self-evolving latent reasoning transfer for generative recommendation.


\newpage
\bibliographystyle{ACM-Reference-Format}
\bibliography{ref}

\newpage
\appendix
\section{Additional Experimental Analysis}
\subsection{Cross-Domain Generalization}
\begin{table}[t]
\centering
\caption{Cross-dataset generalization results on Beauty and Sports.}
\label{tab:cross_dataset}
\setlength{\tabcolsep}{5pt}
\begin{tabular}{lcccc}
\toprule
\textbf{Model} 
& \textbf{Beauty-Sports} 
& \textbf{Sports-Beauty} \\
\midrule
LASAR 
& 0.0472
& 0.0508 \\
EvoReason 
& 0.0584 
& 0.0635 \\
\bottomrule
\end{tabular}
\end{table}
EvoReason provides more effective and dense semantic reasoning supervision for implicit reasoning, enabled by its self-evolving pseudo-tool library that constrains and structures reasoning trajectories. These advantages allow EvoReason to outperform conventional off-policy distillation methods. As shown in the Tabel~\ref{tab:cross_dataset}, EvoReason exhibits a smaller performance degradation on out-of-domain datasets compared with LASAR, further demonstrating its stronger reasoning capability and better generalization ability.

\begin{table}[t]
\centering
\caption{Inference latency comparison on two public datasets.}
\begin{tabular}{llccc}
\toprule
Dataset & Method & Time/Sample (s) & Total Time \\
\midrule
\multirow{3}{*}{Beauty}
& Tiger & 0.53 & 26min \\
& EvoReason & 0.56 & 27min \\
& Onerec-Think & 6.7 & 4.5h \\
\midrule
\multirow{3}{*}{Sports}
& Tiger & 0.54 & 49min \\
& EvoReason & 0.55 & 51min \\
& Onerec-Think & 6.7 & 7.9h \\
\bottomrule
\end{tabular}
\label{tab:latency}
\end{table}
\subsection{Inference Efficiency}
Table~\ref{tab:latency} summarizes the inference latency on two public datasets under the same experimental settings as described in our implementation details. Compared with TIGER, OneReason introduces only a marginal increase in latency due to the additional computation of implicit reasoning tokens in generative recommendation. In contrast, OneRec-Think incurs substantially higher latency because it performs explicit CoT reasoning before generating the final answer, resulting in significantly increased computational overhead.

\begin{table}[t]
\centering
\caption{Sensitivity analysis of the primitive library size $K$ on Beauty and Sports datasets.}
\label{tab:primitive_size}
\begin{tabular}{lccc}
\toprule
Dataset & $K=3$ & $K=5$ (default) & $K=10$ \\
\midrule
Beauty  & 0.0703 & 0.0724 & 0.0718 \\
Sports  & 0.0623 & 0.0657 & 0.0642 \\
\bottomrule
\end{tabular}
\end{table}

\subsection{Parameter Sensitivity Analysis}
The size of the self-evolving primitive library plays an important role in CoT generation, where the library size $K$ is treated as a key hyperparameter. In our experiments, we set $K=5$ by default. To investigate the impact of the primitive library size, we conduct a sensitivity analysis by varying $K$ among $\{3,5,10\}$. As shown in Table~\ref{tab:primitive_size}, the model achieves the best performance when $K=5$, demonstrating that an appropriately sized primitive library provides more effective guidance for CoT generation. We observe that a smaller library size ($K=3$) introduces insufficient CoT constraints, leading to weaker reasoning guidance and degraded performance. In contrast, an overly large library size ($K=10$) introduces additional primitive tools that may bring redundant or misleading signals, interfering with the CoT generation process. Therefore, $K=5$ achieves a better balance between sufficient reasoning guidance and avoiding unnecessary interference.



\end{document}